\documentclass[a4paper]{article}

\usepackage{amsfonts}

\def\be{\begin{equation}}
\def\ee{\end{equation}}
\def\bea{\begin{eqnarray}}
\def\eea{\end{eqnarray}}
\def\({\left(}
\def\){\right)}
\def\<{\left<}
\def\>{\right>}

\def\[{\left[}
\def\]{\right]}
\def\tr{{\mbox{tr}}}
\def\be{\begin{equation}}
\def\ee{\end{equation}}
\def\bea{\begin{eqnarray}}
\def\eea{\end{eqnarray}}
\def\({\left(}
\def\){\right)}
\def\<{\left<}
\def\>{\right>}

\def\[{\left[}
\def\]{\right]}
\def\+{\bar}
\def\mb{\mathbb}
\def\tr{{\mbox{tr}}}
\begin{document}

\pagestyle{empty}
\vskip-10pt
\vskip-10pt
\hfill 
\begin{center}
\vskip 3truecm
{\Large \bf
Loop space, $(2,0)$ theory, and solitonic strings}\\ 
\vskip 2truecm
{\large \bf
Andreas Gustavsson}\footnote{a.r.gustavsson@swipnet.se}\\
\vskip 1truecm
{\it F\"{o}rstamajgatan 24,\\
S-415 10 
G\"{o}teborg, Sweden}\\
\end{center}
\vskip 2truecm
{\abstract{We present an interacting action that lives in loop space, and we argue that this is a generalization of the theory for a free tensor multiplet. From this action we derive the Bogomolnyi equation corresponding to solitonic strings. Using the Hopf map, we find a correspondence between BPS strings and BPS monopoles in four-dimensional super Yang-Mills theory. This enable us to find explicit BPS saturated solitonic string solutions.}}

\vfill 
\vskip4pt
\eject
\pagestyle{plain}

\section{Introduction}
In this paper we will investigate the interacting generalization of the free tensor multiplet theory in six dimensions. These theories have $(2,0)$ supersymmetry and are called $(2,0)$ theories. Since there is a two-form gauge potential in the Abelian tensor multiplet, one might think that the interacting generalization would involve some kind of `non-Abelian two-form'. During the years many people have contributed to the construction of a non-Abelian two-form, which in addition requires the introduction of a one-form and a certain flatness condition (see \cite{Alvarez} and \cite{Schreiber} and references therein). Non-Abelian surfaces holonomies can be defined using this formalism. But as far as I know, no `non-Abelian' action that reduces to the Yang-Mills action upon compactification on a circle, has been possible to construct in this formalism. Another suggestion has been that the non-Abelian two-form should carry three gauge indices rather than the usual two. This has also been successfully used to construct non-Abelian surface holonomies \cite{Akhmedov}. But again the action for such two-forms became just a copy of Abelian actions, which we think is highly dissatisfactory if it were to describe $(2,0)$ theory. We therefore think that there are now good reasons to doubt that one would be able to use any kind of local non-Abelian two-form as dynamical variable of $(2,0)$ theory.

Our approach to $(2,0)$ theory is inspired by the fact that an Abelian gerbe on a manifold $M$ is equivalent with a line bundle over loop space $LM$. Loop space is defined as the space of mappings $S^{1}\rightarrow M$. A derivation of this fact, which uses Cech cohomology, can be found in \cite{Hitchin}. Gerbes arise whenever one has a higher rank gauge field. In the Abelian tensor multiplet we have an anti self-dual two-form gauge field $B$, anti self-dual in the sense that its field strength $H=dB$ is anti self-dual, $H=-*H$. Though when quantizing this theory, one should include the self-dual part of $H$ as well. This will not be part of the tensor multiplet and it does not couple to anything else. One then carries out holomorphic factorization to obtain correlation functions in one of the chiral theories \cite{Witten:1996hc,Henningson:1999dm}. 

The connection two-form can be viewed as a connection one-form on the line bundle over $LM$. The one-form has been given as 
\bea
A_{\mu s}(C)=B_{\mu\nu}(C(s))\dot{C}^{\nu}(s)\label{ab}
\eea
in \cite{Nepomechie,Schreiber}. Here $\mu$ is a space-time vector index, whereas the entity $(\mu s)$ is a loop space vector index. $C$ denotes a point in loop space and is represented as a parametrized loop $s\mapsto C^{\mu}(s)$ in space-time, with tangent vector $\dot{C}^{\mu}(s)$. 

We will not assume that Eq (\ref{ab}) holds in this paper. Instead we will take $A_{\mu s}(C)$ to be the fundamental field, which in general depends on the entire loop $C$ in a non-local way. We will promote $A_{\mu s}(C)$ to a dynamical variable. For one thing, there seems to be no other way in which we could get the equations of motions that we want (and which were derived in \cite{AG2} from $(2,0)$ supersymmetry alone) from an action in loop space, but to let $A_{\mu s}(C)$ be a dynamical variable that we vary in order to derive the equations of motion. We will use the name `loop field' for such a field that lives on loop space. 

If anything we do in loop space is supposed to have any application in physics (which of course is what we hope!), then the most basic requirement ought to be that we can recover the theory for the usual local Abelian two-form gauge field. In section \ref{abelian} we will construct a two-form $b_{\mu\nu}(x)$ out of $A_{\mu s}(C)$ and argue that this two-form is a  gauge field with the expected dynamical quantum behavior, though our argument is far from being complete. In particular we will not address any global issues, such as how this gauge field should transform between two overlapping patches of the loop space manifold.

We will take the point of view that the loops are the fundamental objects in the theory which are created and annihilated by the quantum loop fields. It is tempting to interpret the loops as tensionless selfdual strings. However we will not make any attempt to study their dynamics in this paper. This should be a difficult problem since the loops (strings) are strongly coupled. Due to self-duality and a Dirac type of charge quantization condition for self-dual strings in six dimensions, the coupling constant in $(2,0)$ theory is a fixed number of order unity \cite{Witten} which can never be made small. This means that these interacting quantum theories can not be analysed perturbatively by starting with some classical action. But even so, a classical action can be used to derive classical solitonic solutions, and, perhaps, also to study the quantum theory for the fluctuations around such classical solutions by expanding the quantum loop fields about such a classical field configuration and only treat the fluctuations quantum mechanically. 

In section \ref{s} we introduce the notion of general covariance in loop space. Since our loop space consists of parametrized loops, reparametrization invariance is nothing but a special case of general covariance in loop space. In section \ref{susyaction} we introduce our loop space notations and present the classical action for the `non-Abelian tensor multiplet' in loop space. In section \ref{sectcanonical} we compute the anti-commutator of two supercharges, and obtain the central charges that correspond to a self-dual string and a three brane respectively. We then obtain the Bogomolnyi equation for strings. In section \ref{implications} we use the Hopf map to project the Bogomolnyi equation to the familiar Bogomolnyi equation of super Yang-Mills theory. In section \ref{bps} we use this result to find explicit BPS string solitons. In section \ref{abelian} we scetch how one might be able to recover the Abelian theory for a local two-form gauge field from the loop space theory for our non-local Abelian loop field $A_{\mu s}(C)$.

\section{Reparametrization invariance from general covariance}\label{s}
A motivation for Eq (\ref{ab}) comes from identifying the Wilson surface\footnote{Here $(t,s)\mapsto X^{\mu}(t,s)$ is the embedding map of the surface, $\cdot\equiv d/dt$ and $' \equiv d/ds$}
\bea
\int dt\int ds B_{\mu\nu}(X(t,s))\dot{X}^{\mu}(t,s)X'^{\nu}(t,s)
\eea
with the Wilson loop
\bea
\int dt A_{\mu s}(C(t))\dot{C}^{\mu s}(t)
\eea
in loop space, where we define
\bea
C^{\mu s}(t)\equiv X^{\mu}(s,t).
\eea
and we use the Einstein summation convention $V_{\mu s}U^{\mu s} \equiv \sum_{\mu}\int ds V_{\mu s}U^{\mu s}$.

Both these observables should be diffeomorphism invariant (modulo $2\pi$). The condition for the former to be invariant is that $B_{\mu\nu}$ transforms as a tensor of rank two. The condition for the latter to be invariant should be that $A_{\mu s}$ transforms as a vector or, in other words, as one-form, in loop space -- whatever that means. 

We will formulate our theory in loop space. To this end we drop everything that separates out space-time as anything particular. So we drop the constraint (\ref{ab}) on our one-form and will not assume that $A_{\mu s}(C)$ has been constructed out of a local two-form. But we will assume that $A_{\mu s}$ transforms as a vector. To make this statement precise, we introduce the notation
\bea
\partial_{\mu s} = \frac{\delta}{\delta C^{\mu}(s)}
\eea
for the usual functional derivative. We then define a vector in loop space to be a quantity that transforms as
\bea
\delta A_{\mu s} = (\partial_{\mu s} \epsilon^{\nu t})A_{\nu t} + \epsilon^{\nu t}\partial_{\nu t} A_{\mu s}  \label{Eq}
\eea
under the infinitesimal diffeomorphism 
\bea
\delta C^{\mu s}(C) = - \epsilon^{\mu s}(C).
\eea
Among these diffeomorphism we find those induced by space-time diffeomorphisms as  
\bea
\epsilon^{\mu s}(C)=\epsilon^{\mu}(C(s))
\eea
and those induced by reparametrizations as
\bea
\epsilon^{\mu s}(C)=\epsilon(s)\dot{C}^{\mu}(s)
\eea
One reason for wanting to abandon Eq (\ref{ab}) where $A_{\mu s}$ is defined in terms of a two-form, is that this $A_{\mu s}$ does not quite transform as a vector (in the sense of Eq (\ref{Eq})) under diffeomorphisms in loop space\footnote{It transforms as $\delta A_{\mu s} = \epsilon^{\rho}(C(s))\partial_{\rho} A_{\mu s} + (\partial_{\mu}\epsilon^{\rho}(C(s)))A_{\rho s}$.}, not even under diffeomorphisms induced by space-time diffeormorphisms, $\epsilon^{\mu}(C(s))$. We can consider much more general diffeomorphisms in loop space though. For instance $\epsilon^{\mu s}(C) = \epsilon^{\mu}(s\partial_s{C}(s))$ or whatever. We would like to contruct a theory that is covariant under any such a general diffeormorphism in loop space. That would give us reparametrization invariance (and of course space-time diffeomorphism invariance) for free. 

If we insert that $\epsilon^{\rho r}=-\epsilon(r)\dot{C}^{\rho}(r)$ in Eq (\ref{Eq}), we can derive that $A_{\mu s}$ transforms as a one-form under reparametrizations,\footnote{Notice that $\delta s=-\epsilon$ $\Leftrightarrow$ $\delta C^{\mu}(s)=\epsilon(s)\dot{C}^{\mu}(s)$ which explains the relative minus sign.}
\bea
\delta A_{\mu s} &=& -\int dr\(\epsilon(r)\dot{C}^{\rho}(r)\partial_{\rho r}A_{\mu s} - \partial_{\mu s}(\epsilon(r)\dot{C}^{\rho}(r))A_{\rho r}\)\cr
&=& -\int dr \(\epsilon(r)\partial_r A_{\mu s} - \epsilon(r)\dot{\delta}(r-s)A_{\rho r}\)\cr
&=& (\partial_s\epsilon(s))A_{\mu s} + \epsilon(s)\partial_s A_{\mu s}
\eea
One should be confused about this derivation since $\partial_{\mu s}A_{\nu t}$ is expected to be very singular at $s=t$. But $\partial_s A_{\nu t}$ on the other hand should not be singular anywhere since here $\partial_s$ denotes the usual derivative with respect to the parameter $s$ and we expect $A_{\mu s}$ to be a smooth function of $s$. So $\partial_s A_{\nu t}$ should vanish almost everywhere (except when $s=t$ where we expect to get a finite value.) So apparently when we contract $\dot{C}^{\mu}(s)$ with $\partial_{\mu s}A_{\nu t}$ we must kill the singular behaviour at $s=t$ (which means that $\dot{C}^{\mu}(s)$ must be orthogonal to those divergencent directions. This one may check explicitly for the case that $A_{\mu s}$ is given as in Eq (\ref{ab}):  $\int ds \dot{C}^{\mu}(s)\partial_{\mu s}A_{\nu t} = \dot{C}^{\mu}(s)\dot{C}^{\rho}(s)\(\partial_{\mu}B_{\nu\rho}-\partial_{\rho}B_{\nu\mu}\)\equiv 0$).

We define a contra-variant vector as a vector that transforms according to
\bea
\delta V^{\mu s}=\epsilon^{\rho r}\partial_{\rho r}V^{\mu s} - (\partial_{\rho r}\epsilon^{\mu s})V^{\rho r}.\label{contra}
\eea
We can now build a scalar by contracting a co-variant vector with a contra-variant as $U_{\mu s}V^{\mu s}$. Since the index $s$ is supposed to be contracted this means that this quantity in particular should be reparametrization invariant (and not only covariant). But if we just plug in $\epsilon^{\rho r}=-\epsilon(r)\dot{C}^{\rho}(r)$ in Eq (\ref{contra}) we get, after a similar computation as we did for the co-variant vector, that
\bea
\delta V^{\mu s}=\epsilon(s)\partial_s V^{\mu s}\label{sc}
\eea
(Here we do not sum over $s$). But this is not the transformation rule that one would expect of a contra-variant vector under reparametrizations! The thing is that we also have to take into account the transformation property of $ds$. In order for $U_{\mu s}V^{\mu s}\equiv \int ds U_{\mu s}V^{\mu s}$ to be invariant, $U_{\mu s}V^{\mu s}$ should not be a scalar but a co-vector. Or in other words, $V^{\mu s}$ should be a scalar (and not a contra-variant vector) if $U_{\mu s}$ is a co-vector. And this is indeed precisely what we found in Eq (\ref{sc}). 

It is natural to associate one differential $ds$ with each rised vector index. Hence we will let $V^{\mu s}=ds V^{\mu s}$. Anyhow we need one $ds$ in order to perform the contraction of the $s$-indices. Then all tensors will transform as one would expect from the position of their indices. For instance $U^{ss's''}$ will transform as 
\bea
\delta U^{ss's''} &=& \(\epsilon(s)\partial_s + \epsilon(s')\partial_{s'} + \epsilon(s'')\partial_{s''}\)U^{ss's''}\cr
&&-\(\partial_s\epsilon(s)+\partial_{s'}\epsilon(s')+\partial_{s''}\epsilon(s'')\)U^{ss's''}
\eea
and so on.
 
To be able to construct a generally covariant theory we should have a metric tensor $G_{\mu s,\nu t}$ in loop space, that tranforms as a tensor of rank two,
\bea
\delta G_{\mu s,\nu t} = (\partial_{\mu s}\epsilon^{\rho r}) G_{\rho r,\nu t} + (\partial_{\nu t} \epsilon^{\rho r}) G_{\mu s,\rho r} + \epsilon^{\rho r}\partial_{\rho r}G_{\mu s,\nu t}.
\eea
Restriciting to reparametrizations, $\epsilon^{\mu s}=-\epsilon(s)\dot{C}^{\mu}(s)$, we get
\bea
\delta G_{\mu s,\nu t} &=& \epsilon(s)\partial_s G_{\mu s,\nu t} + \epsilon(t)\partial_t G_{\mu s,\nu t} \cr
&& + (\partial_s \epsilon(s)) G_{\mu s,\nu t} + (\partial_t\epsilon(t))G_{\mu s,\nu t}\label{Eqw}
\eea
We will choose the metric to be of the form
\bea
G_{\mu s,\nu t}=g_{\mu\nu}(C(s)) \delta(s-t)\label{metri}
\eea
Unfortunatley it is not manifest that this metric transforms in a covariant way under reparametrizations. From Eq (\ref{Eqw}) we find that
it should transform as 
\bea
\delta G_{\mu s,\nu t} = (\partial_s\epsilon(s)) G_{\mu s,\nu t}.
\eea
This would have been so if $\delta(s-t)$ had transformed as a scalar
\bea
\delta(\delta(s-t)) = (\epsilon(s)\partial_s + \epsilon(t)\partial_t)\delta(s-t).
\eea
but this is not how the delta function really transforms. It does not transform at all,
\bea
\delta(\delta (s-t))=0
\eea
But it is not hard to find a quantity $d(s',t')$ with the desired transformation property and which is such that $d(s,t)=\delta(s-t)$. One may take 
\bea
d(s',t')=\frac{ds}{ds'}\delta(s-t).
\eea
The manifestly covariant metric is then $G_{\mu s,\nu t}=g_{\mu\nu}(C(s))d(s,t)$.

We will use $A_{\mu s}(C)$ to denote also a non-Abelian gauge loop field. We take the point of view that a Wilson surface is nothing but a Wilson line in loop space. Hence we define the Wilson surface just as one defines a Wilson line,
\bea
W(\Gamma)=\tr P \exp \int_{\Gamma} dC^{\mu s}A_{\mu s}
\eea
where $\Gamma$ denotes a line in loop space, and $P$ denotes path ordering along this line. 

In the Introduction we said that the coupling constant in $(2,0)$ theory is a fixed number. We will now argue that we have infinitely many coupling constants $g^s$, one for each value of $s$, and which constitute a  contra-variant vector with respect to reparametrizations of $s$. We can then bring all these coupling constants into any fixed value by a reparametrization. Taking global issues into account, we expect that it will turn out to be so that one can bring them all into one and only one fixed number, which is equal to the (invariant) coupling constant of $(2,0)$ theory. In the generic situation we define the Wilson surface as
\bea
W(\Gamma) = \tr P \exp \int dt \int ds g^s\frac{dC^{\mu s}(t)}{dt}A_{\mu s}(C(t))
\eea
and we will now examine the conditions of reparametrization invariance. This will essentially be a repetition of what we said in \cite{AG1}, but since the introduction of the coupling constants $g^s$ is new, let us repeat the arguments in all its details.

Given a surface $\Sigma$ embedded in space-time, there are many different lines $\Gamma$ we could consider. If we refer to the parameters $s^A\equiv (t,s)$ of the surface as time and space coordinates, then different $\Gamma$'s correspond to different ways of choosing constant time slicings of the surface. On a generic surface there is no distinguished time slicing that can be used to define the Wilson surface. So in order to get a well-defined Wilson surface we have to ensure that we get the same value irrespectively of which constant time slicing we use. 

We thus associate $t$ with time, and the loop associated with a constant time $t$ we denote as
\bea
C^{A}_t(s)\equiv X^A(t,s)=s^A.
\eea
(This is the loop $C$ expressed in the parameter space of the surface.) We are then interested in deforming such loops. We should not really assume that the $C^{A}_t(s)$ are straight lines to start with as this is a rather degenerate situation. So we assume that we have already made an arbitrary reparametrization $(t,s)\rightarrow (t',s')$. Another way to express this is to say that the loops $C^A_t(s)$ need not be of the straight lines $C^A_t(s)=(t,s)$. We then make an infinitesimal variation of such loops and require the Wilson surface to be invariant under such a variation.

We let 
\bea
A_{A s}=\partial_A X^{\mu}(t,s) A_{\mu s}
\eea
be the pullback of the one-form connection to the surface. If we then associate 
\bea
H(t)=\int ds \frac{dC^{A s}(t)}{dt} A_{A s}(C(t)).
\eea
with the Hamiltonian, then $W$ will be the generator of time translation and we can apply the formalism of generalised Hamiltonian dynamics \cite{Teitelboim}.

Given a loop, there is a distinguished set of tangent vectors to the surface along this loop, namely $\partial_s C^{A}(s)$ and $n^A(s)$ where $n^A(s)$ is the unit normal vector to the loop,
\bea
g_{AB}(C(s))n^A(s) n^B(s) &=& 1,\cr
g_{AB}(C(s))n^A(s) \partial_s C^B(s) &=& 0
\eea
and $g_{AB}$ denotes the induced metric on the surface. Generically the Hamiltonian is written as
\bea
H(t) = \int ds \frac{\partial C^{A}_t(s)}{\partial t} {\cal{H}}_A
\eea
One then projects on the normal $\perp$ and tangential $s$ directions as
\bea
{\cal{H}}^A = n^A {\cal{H}}_{\perp} + (\partial^s C^A) {\cal{H}}_s
\eea
where the induced metric on the loop, 
\bea
\gamma(s)=\partial_s C^A(s)\partial_s C_A(s)
\eea
is used to rise the index $s$ as 
\bea
\partial^s C = \gamma^{-1}\partial_s C
\eea
The result in \cite{Teitelboim} is that the time evolution operator is independent of the choice of fibration $C^{\mu}_t(s)$ if and only if these Hamiltonian components obey certain commutation relatations (or Poisson bracket relations in a classical Hamilton theory). 

For our application of the Wilson surface, we find that
\bea
{\cal{H}}_{\perp} &=& g^s n^A(s) A_{A s} \cr
{\cal{H}}_s &=& g^s (\partial_s C^A)A_{A s}
\eea
We now have to ensure that the equations for reparametrization invariance are satisfied. These equations imply complicated conditions for the commutator $[A_{A s},A_{B t}]$, unless we impose 
\bea
{\cal{H}}_s &\sim & (\partial_s C^A)A_{A s} \equiv 0
\eea
as a strong constraint, that has to vanish also inside commutators (or Poisson brackets). Then the conditions for reparametrization invariance become very simple, namely just
\bea
[{\cal{H}}_{\perp}(s),{\cal{H}}_{\perp}(t)] = 0.
\eea

We can satisfy this constraint by letting
\bea
A_{\mu s} = A^a_{\mu s}\lambda_a(s)
\eea
where 
\bea
[\lambda_a(s),\lambda_b(t)] = C_{ab}{}^c(s) \delta(s-t) \lambda_c(s)
\eea
and the $C_{ab}{}^c(s)$ are structure constants of a Lie group $G_s$. The gauge group is the infinite tensor product 
\bea
\bigotimes_s G_s
\eea
Now since a reparametrization in $s$ moves the different $G_s$'s around, reparametrization invariance forces us to take all the $G_s$ to be different copies of one and the same group (for instance each $G_s=SU(N)$). But we still have the possibility to have different coupling constants $g^s$ in each $G_s$ factor. If we let $C_{ab}{}^c(s)$ be the same for each $s$ then we have to put the coupling constants in the Wilson surface and in the definition of the fields strength. The generators $\lambda^a(s)$ will then transform as co-variant vectors, which makes integrals like $\int ds g^s A^a_{\mu s}\lambda_a(s)$ or $\int ds g^s dC^{\mu s}A^a_{\mu s}\lambda_a(s)$ well-defined (i.e. invariant under reparametrizations of $s$). The gauge field strength should then be defined as
\bea
F_{\mu s,\nu t}=\partial_{\mu s}A_{\nu t}-\partial_{\nu t}A_{\mu s} + g^s[A_{\mu s},A_{\nu t}].
\eea
(with no sum over $s$).

We can also put the coupling constants in the structure constants by defining new generators as
\bea
g^s\lambda^a(s)=\lambda'^a(s)
\eea
which then will transform as scalars, and will obey the algebra
\bea
[\lambda_a(s),\lambda_b(t)] = C_{ab}{}^c g^s \delta(s-t) \lambda_c(s).
\eea
We then let $g^s A^a_{\mu s}\lambda^a(s)=A^a_{\mu s}\lambda'^a(s) \equiv A'_{\mu s}$ and the Wilson surface and gauge field strength become
\bea
W(\Gamma) &=& \tr P \exp \int dt \int ds \frac{dC^{\mu s}(t)}{dt}A_{\mu s}(C(t)),\cr
F_{\mu s,\nu t} &=& \partial_{\mu s}A_{\nu t}-\partial_{\nu t}A_{\mu s} + [A_{\mu s},A_{\nu t}].
\eea

On the gauge field $A_{\mu s}$ we still have to impose the constraints
\bea
\dot{C}^{\mu}(s)A_{\mu s}(C)=0.
\eea
One could use Lagrange multiplicators in the action to implement these constraints. That action will be singular and hence generate secondary constraints and one finds a reduced phase space on which one should use the Dirac brackets. But we have not performed this analyzis in detail. 

Let us notice that the constraints are metric independent. The tangent vector $\dot{C}^{\mu}(s)$ is born with its vector index up-stairs while $A_{\mu s}$ is born with its co-vector index down-stairs, so these objects can be contracted without using the metric. The two constraints found above are gauge invariant at least under local gauge transformations. Under such a gauge transformation 
\bea
\delta A_{\mu s}(C)=\frac{\delta\lambda(C)}{\delta C^{\mu}(s)}+[A_{\mu s}(C),\lambda(C)]
\eea
the second constraint transforms by
\bea
\dot{C}^{\mu}(s)\frac{\delta\lambda(C)}{\delta C^{\mu}(s)}+[\dot{C}^{\mu}(s)A_{\mu s}(C),\lambda(C)]=\frac{d\lambda(C)}{ds}
\eea
Thus if we require the gauge parameter $\lambda(C)$ be subject to the condition that $d\lambda/ds=0$, then the second constraint is gauge invariant. The constraint $A_{\mu s}\dot{C}^{\mu}(s)=0$ for any fixed $s$ is of course not reparametrization invariant. Making an infinitesimal reparametrization we generate the constraint $\partial_s(A_{\mu s}\dot{C}^{\mu}(s))=0$, and we may repeat this procedure to produce constraints $(\partial_s)^n(A_{\mu s}\dot{C}^{\mu}(s))=0$ for any number $n$ of derivatives. This means that we must take $A_{\mu s}\dot{C}^{\mu}(s)=0$ for all $s$, and this infinite set of constraints are closed under reparametrizations.

\section{The $(2,0)$ supersymmetric action in loop space}\label{susyaction}
We will assume a flat Minkowski space $M={\mb R}^{1,5}$, with metric tensor $\eta_{\mu\nu}=$ diag$(-1,1,1,1,1,1)$. As coordinates $C^{\mu s}$ in loop space $LM$ we take the set of mappings $s\mapsto C^{\mu}(s)\equiv C^{\mu s}$ where $C^{\mu}(s)=C^{\mu}(s+2\pi)$, and we will take $s\in [0,2\pi]$. We will assume that our loop space comes equipped with the metric 
\bea
G_{\mu s,\nu t}=\eta_{\mu\nu}2\pi \delta(s-t).
\eea 
We will also assume that we have brought all the coupling constants $g^s=1$ by a reparametrization (and hence we loose manifest reparametrization invariance).

Following \cite{AG2}, we introduce the `non-Abelian tensor multiplet fields' $\phi_{\mu s}(C)$, $A_{\mu s}(C)$ and $\psi_{\mu s}(C)$. These are loop fields that has to be subject to the following constraints
\bea
D^{\mu s}\phi_{\mu s}&=&0\cr
D^{\mu s}\psi_{\mu s}&=&0\cr
A_{\mu s}\dot{C}^{\mu}(s)&=&0\cr
\psi^{[\mu s}\dot{C}^{\nu]}(s)&=&0\cr
\phi^{[\mu s}\dot{C}^{\nu]}(s)&=&0\cr
\phi^A_{[\mu s}\phi^B_{\nu] s}&=&0\cr
\psi_{[\mu s}\phi^A_{\nu]s}&=&0\cr
F_{\mu s,\nu t}\phi^{A,\nu t}&=&0\cr
F_{\mu s,\nu t}\psi^{\nu t}&=&0\label{constr}
\eea
in order for supersymmetry to close on-shell \cite{AG2}. Here 
\bea
D_{\mu s}:=\partial_{\mu s}+A_{\mu s}
\eea
is the gauge covariant derivative with $\partial_{\mu s}=\frac{\delta}{\delta C^{\mu}(s)}$ being the usual functional derivative, which we for later convenience will normalize with a factor of $2\pi$ as
\bea
\int \frac{ds}{2\pi} \frac{\delta f(C)}{\delta C^{\nu}(s)}\delta C^{\nu}(s)=\delta f(C).
\eea
The gauge field strength is given by
\bea
F_{\mu s,\nu t}=\partial_{\mu s}A_{\nu t}-\partial_{\nu t}A_{\mu s}+[A_{\mu s},A_{\nu t}]
\eea
We can also define the spinor loop field
\bea
\psi_s:=\Gamma_{\mu}\psi^{\mu s}.
\eea
for which we find that $\bar{\psi}_s=-\bar{\psi}^{\rho s}\Gamma_{\rho}$ and that this spinor is subject to the symplectic Majorana condition $\bar{\psi}_s=\psi^T_s C$ where $C$ is the eleven-dimensional charge conjugation matrix. We use the same spinor conventions and notations as in \cite{AG2}. 

We will represent all the loop fields (collectively denoted as $\varphi$) as 
\bea
\varphi_{\mu s}=\varphi_{\mu s}^a\lambda^a(s)
\eea
where we take the generators $\lambda^a(s)$ to obey the loop algebra
\bea
[\lambda^a(s),\lambda^b(t)]=C^{abc}2\pi\delta(s-t)\lambda^c(s)
\eea
and normalize them to unity,
\bea
\tr\(\lambda^a(s)\lambda^b(t)\)=\delta^{ab}2\pi \delta(s-t).
\eea
We should assume that $\delta(s-t)$ transforms as a scalar. The generators $\lambda^a(s)$ will then transform as scalars. $\phi_{\mu s}$ will transform as a vector.

We will assume that all the loops can be representented in terms of Fourier modes as\footnote{If there are some compact dimensions we should of course also include winding modes.} 
\bea
C^{\mu}(s)=x^{\mu}+\sum_{m\neq 0} \alpha^{\mu}_me^{ims}
\eea
We define the Fourier modes
\bea
\varphi^{\mu}_m&=&\int \frac{ds}{2\pi} e^{-ims}\varphi^{\mu s}\cr
\alpha^{\mu}_m&=&\int \frac{ds}{2\pi} e^{-ims}C^{\mu}(s)\cr
\partial_{\mu m}&=&\int \frac{ds}{2\pi} e^{-ims}\partial_{\mu s}
\eea
and find that
\bea
\partial_{\mu m}=\frac{\partial}{\partial \alpha^{\mu m}}
\eea
The metric becomes
\bea
G_{\mu m,\nu n}=\eta_{\mu\nu}\delta_{m+n}
\eea
so that $a.b:=a_{\mu m}b^{\mu m}=\sum_m a_{\mu m}b^{\mu}_{-m}$. We also define
\bea
\lambda^a_m:=\int \frac{ds}{2\pi} e^{-ims}\lambda^a(s)
\eea 
which obey the algebra
\bea
[\lambda^a_m,\lambda^b_n]=C^{abc}\lambda_{m+n}^c
\eea
The algebra for $t^a:=\lambda^a_0$ 
\bea
[t^a,t^b]=C^{abc}t^c
\eea
will be identified with the algebra associated with the gauge group. In terms of these modes the loop fields get represented as
\bea
\varphi_{\mu m}=\sum_n \varphi_{\mu n}^a\lambda^a_{m-n}
\eea
We will also need the trace 
\bea
\tr(\lambda^a_m\lambda^b_n) = \delta_{ab}\delta_{m+n}
\eea
We have thus rescaled the generators so that in particular
\bea
\tr(t^at^b)=\delta^{ab},
\eea
which is always possible for simply laced Lie groups. 

We define an average over loops as
\bea
\<...\>:=\int D_x C  e^{-\frac{L^2(C)}{\Lambda^2}}...
\eea
Here the functional integral is over all closed loops centered at the space-time point $x$, $\Lambda$ is a cut-off, and we define a length 
\bea
L^2(C)=\int \frac{ds}{2\pi} \dot{C}_{\mu}(s)\dot{C}^{\mu}(s)=\sum_m m^2\alpha_{\mu m}\alpha^{\mu m}
\eea
In terms of modes we get
\bea
\<...\>_{\alpha;\Lambda}=\int \[\prod_{m\neq 0} d^6\alpha_m\] e^{-\frac{1}{\Lambda^2}\sum_n \alpha_n.\alpha_{-n} n^2}.
\eea
It is an obvious advantage to work with the Fourier modes $\alpha_m^{\mu}$ in place of $C^{\mu}(s)$. While the $C^{\mu}(s)$ are indexed by $s$ which takes values in the uncountable set of real numbers and are also constrained by periodicity, the Fourier modes are labeled by $n$ which takes values in the countable set of integer numbers and are subject to no periodicity constraint.

We now consider the following supersymmetric action 
\bea
S&=&\lim_{\Lambda\rightarrow \infty} N_{\Lambda}\int d^6 x \<{\cal{L}}\>,
\eea
\bea
{\cal{L}}&=&\tr \bigg(\frac{1}{4}F_{\mu m, \nu n}F^{\mu m,\nu n} + \frac{1}{2}D_{\mu m}\phi^A_{\nu n}D^{\mu m} \phi^{\nu n}_A + \frac{1}{4}[\phi^A_{\mu m},\phi^{B,\mu m}]^2 \cr
&&+ \frac{i}{2} \sum_n \int \frac{ds}{2\pi} e^{ins}\Big(\bar{\psi}^m\Gamma^{\mu}D_{\mu n}\psi_m + \bar{\psi}^m\Gamma^{\mu}\Gamma_A[\phi^A_{\mu (n},\psi_{m)}]\cr
&&+\frac{1}{2}\bar{\psi}^mD^{\mu}_{(n}\psi_{\mu m)}\Big)\bigg)\label{action}
\eea
The normalization constant $N_{\Lambda}$ of course depends on the precise way in which we implement the cut-off. Using the above cut-off prescription we will find that
\bea
N_{\Lambda}=\frac{8}{\Lambda^2\<1\>}.
\eea
We notice that we can not do integrations by parts and throw away boundary terms (which will be of $O(\Lambda^{-2})$), unless we take the limit $\Lambda\rightarrow \infty$ and also assume that the fields drop to zero at infinity sufficiently fast. But this is the opposite limit to that which was taken in \cite{Nepomechie}. 

The supersymmetry variations are given by
\bea
\delta \phi^A_{\mu m} &=& -i\bar{\epsilon}\Gamma^A \psi_{\mu m}\cr
\delta \psi_m &=& \sum_n e^{ins}\Big(\frac{1}{2} F_{\mu m,\nu n}\Gamma^{\mu\nu} + D_{\mu m}\phi^A_{\nu n}\Gamma^{\nu\mu}\Gamma_A + \frac{1}{2}[\phi^A_{\mu m},\phi^B_{\nu n}]\eta^{\mu\nu}\Gamma_{AB}\cr
&&-\frac{1}{2}D^{\mu}_{(m}\phi^A_{\mu n)}\Gamma^A\Big)\epsilon\cr
\delta F_{\mu m,\nu n} &=& 2i\bar{\epsilon}\Gamma_{\kappa[\mu}D_{\nu] n}\psi^{\kappa}_m
\eea
These supersymmetry variations close on-shell \cite{AG2} in the sense that 
\bea
[\delta_{\epsilon},\delta_{\eta}]=2i\bar{\epsilon}\Gamma^{\mu}\eta \sum_n e^{ins} \partial_{\mu n}.\label{closure}
\eea
The associated supercurrents are given by
\bea
J_{\mu m}=iN_{\Lambda}\(-\frac{1}{2}F_{\kappa n,\tau m} \Gamma^{\kappa \tau} + D_{\kappa n}\phi^A_{\tau m} \Gamma_A \Gamma^{\tau\kappa} - \frac{1}{2}[\phi^A_{\kappa n},\phi^B_{\tau m}]\eta^{\kappa\tau}\Gamma_{AB}\)\Gamma_{\mu} \psi^n.
\eea
To show this it one has to use the constraints in the form
\bea
F_{\mu m,\nu n}\psi^{\mu m}&=&0\cr
\phi_{[\mu |m|}\psi^m_{\rho]}&=&0
\eea
To show that the action is supersymmetric one must use the Bianchi identity
\bea
D_{\mu m}F_{\nu n,\rho p}+D_{\nu n}F_{\rho p,\mu m}+D_{\rho p}F_{\mu m,\nu n}=0
\eea
and make some integration by parts. One also have to use the above constraints. In the variation of the action we also find terms that involve three fermions. All these can be seen to vanish identically by a Fierz rearrangement if one also uses the constraints of the form $\bar{\psi}^{[\mu s}\Gamma_{\nu_1\nu_2\cdots}\psi^{\nu]s}=0$.

The supercurrent is conserved in the sense that $\partial^{\mu m}J_{\mu m}=0$. The conserved (time-independent) supercharges are given by
\bea
Q:=\int d^5 x \<J_{00}\>
\eea
They are time-independent because
\bea
\partial^0 Q=\int d^5 x \<\partial^{0}J_{00}\>=\int d^5 x \<\partial^{\mu m}J_{\mu m}\>=0
\eea
Here one uses that derivatives with respect to oscillators ($m\neq 0$) are total derivatives over which we integrate when we take the average. Hence the result is $O(\Lambda^{-1})$, which is $0$ in the limit $\Lambda\rightarrow\infty$.

\section{Central charges for extended objects}\label{sectcanonical}
We define time in loop space as
\bea
x^0=\int \frac{ds}{2\pi} C^0(s)
\eea
We then find the canonical momenta, 
\bea
E^{im;a}&=&N_{\Lambda}F^{00,im;a}\cr
\pi^a_{im}&=&N_{\Lambda}\(D^0\phi_{im}\)^a\cr
\pi_m^a&=&N_{\Lambda}\frac{i}{2}\bar{\psi}^a_m \Gamma^0
\eea
conjugate to $A_{im}^a,\phi_{\mu m}^a, \psi_m^a$ respectively, and satisfying the equal time canonical commutation relations
\bea
[A^a_{im}(x,\alpha),E^{jn;b}(y,\beta)]&=&\delta^{ab}\delta_{m+n}\delta_i^j\delta^5(x-y)\prod_{n\neq 0}\delta^6(\alpha_n-\beta_n)\cr
[\phi^{a,im}(x,\alpha),\pi^b_{jn}(y,\beta)]&=&\delta^{ab}\delta_{m+n}\delta^i_j\delta^5(x-y)\prod_{n\neq 0}\delta^6(\alpha_n-\beta_n)\cr
\{\psi^a_m(x,\alpha),\bar{\psi}^b_n(y,\beta)\}&=&i\Gamma^0 N^{-1}_\Lambda\delta^{ab}\delta_{m+n}\delta^5(x-y)\prod_{n\neq 0}\delta^6(\alpha_n-\beta_n)
\eea
The `missing' factor of $2$ in the fermionic anti-commutatation relation is due to the fact that the spinors obey a symplectic Majorana condition.

Anti-commuting two supercharges, we get\footnote{When we write $\{Q,\bar{Q}\}$, the anti-commutator acts on the operators only.}
\bea
\{Q,\bar{Q}\}=2\Gamma^0\(\Gamma^{\mu}P_{\mu}+\Gamma^{\mu}\Gamma_A Z_{\mu}^A + \Gamma^{\mu\nu\rho}\Gamma_{AB} W^{AB}_{\mu\nu\rho}\)
\eea
with central charges
\bea
Z_{\mu}^A &=& \frac{N_{\Lambda}}{2}\int d^5 x \Big< \tr\(F_{im,jn}D_k^m\phi^{A,n}_l\)\Big>\epsilon^{ijkl0}{}_{\mu}\cr
W_{\mu\nu\rho}^{AB} &=& \frac{N_{\Lambda}}{2}\int d^5 x \Big<\tr\(D_i\phi_j^A D_k\phi^B_l\)\Big> \delta^{il}\epsilon^{ik0}{}_{\mu\nu\rho}.
\eea
The central charge that corresponds to parallel self-dual strings aligned in the $x^5$ direction is thus given by $Z^A_{\mu}=\delta^A_5 \delta_{\mu}^5 Z$ where
\bea
Z=\frac{N_{\Lambda}}{2}\int d^4 x \epsilon^{ijkl} \<\tr (F_{im,jn}D_k^m\phi_l^n)\>
\eea
if we assume that $\phi^5_{\mu m}$ is the only non-zero scalar loop field. Hence $i,j,..=1,2,3,4$ run over the transverse directions to the strings. Using the Bianchi identity
\bea
D_{[im}F_{jn,kp]}=0
\eea
we can rewrite this central charge as
\bea
Z=\frac{N_{\Lambda}}{2} \int d^4 x  \epsilon^{ijkl} \<\partial_{k}^m\tr (F_{im,jn}\phi_l^n)\> 
\eea
Here only $m=0$ gives a contribution when applying $\<...\>$ as it is otherwise a total derivative terms that give negligible contributions (of order $\Lambda^{-1}$). Hence
\bea
Z=\frac{N_{\Lambda}}{2} \int d^4 x  \epsilon^{ijkl} \<\partial_{k}\tr (F_{i0,jn}\phi_l^n)\>
\eea
This lead us to define
\bea
H_{ijk}(x)a:=\frac{N_{\Lambda}}{2}\<\tr (F_{[i0,jn}\phi_{k]}^n)\>
\eea
since then the central charge can be written in terms of a topological charge as
\bea
Z&=&a\int d^4 x \epsilon^{ijkl}\partial_k H_{ijk}(x)\cr
&=&a \int_{S^3_{\infty}} H 
\eea
if we can identify $a$ with the Higgs scalar vacuum expectation value and $\int H$ with a magnetic charge. Indeed, in the Higgs vacuum where
\bea
D_{\mu m}\phi_{\nu n}&=&0\cr
D^{\mu m}F_{\mu m,\nu n}&=&0\cr
D_{[\mu m}F_{\nu n,\rho p]}&=&0,
\eea
we get
\bea
\partial^i \tr(F_{i0,jn}\phi_k^n)&=&\tr((D^i F_{i0,jn})\phi_k^n)\cr
\partial^i \tr(F_{k0,jn}\phi_i^n)&=&\tr((D^i F_{k0,jn})\phi_i^n)
\eea
Here we use the Bianchi identity on the second equation to get
\bea
(D_i F_{k0,ln})\phi^{in} = -(D_k F_{ln,i})\phi^{in} + (D_{ln} F_{ki})\phi^{in}
\eea
Both terms here vanishes identically by a constraint and $D_{\mu m}\phi_{\nu n}=0$ in the Higgs vacuum. Hence we find that
\bea
\partial^i H_{ijk}=0
\eea
In a similar way one can show that the Biachi identity 
\bea
\partial_{[i}H_{jkl]}=0
\eea
holds in the Higgs vacuum.

\section{Bogomolnyi equations}\label{implications}
The energy is minimized if the Bogomolny equation 
\bea
F_{im,jn}=\pm \epsilon_{ijkl}D^{k}_{(m}\phi^l_{n)}\label{Bog1}
\eea
is satisfied.\footnote{A direct consequence of this equation together with $F_{im,jn}=-F_{jn,im}$ is that $F_{im,jn}=-F_{jm,in}=F_{in,jm}$ and hence we should symmetrize $m,n$ in the right-hand side.} To see this, we rewrite the energy of a static configuration as follows (with our maths conventions we get a minus sign, but that is just because some $i$'s are buried in our fields, so never mind)
\bea
E&=&-N_{\Lambda}\int d^5 x\tr\<\frac{1}{4}F_{im,jn}^2+\frac{1}{2}\(D_{im}\phi_{jn}\)^2+...\>\cr
&\geq&-\frac{N_{\Lambda}}{4}\int d^5 x\tr\<(F_{im,jn}\pm \epsilon_{ijkl}D^k_m\phi^l_n)^2\> \cr
&&\pm \frac{N_{\Lambda}}{2}\int d^5 x\tr\<\epsilon_{ijkl}F^{im,jn}D^{k}_m\phi^{l}_n\>
\eea
Here $...$ involve terms which are $\geq 0$ (like terms that involve $\partial_5$ derivatives). We have used that $\<2D_{\mu m}\phi_{\nu n}D^{[\mu}\phi^{\nu]}\>=\<D_{\mu m}\phi_{\nu n}D^{\mu}\phi^{\nu}\>+O(\Lambda^{-1})$ which can be seen by making an integration by parts and using the constraint $D_{\mu m}\phi^{\mu m}=0$. We see that the BPS bound $E\geq |Z|$ is saturated by field configurations that satisfy the above string Bogomolnyi equation.

A string in five spatial dimensions can be enclosed by an $S^3$. Contrary to $S^2$ (and all other even-dimensional spheres), $S^3$ can be fibrated by loops over $S^2$ (the Hopf fibration). It is curious that precisely in this situation we also have a theory with fields that ought to be evaluated on loops rather than on points. Thinking that this cannot be a mere coincidence, we are led to investigate what we get when we evaluate the loop fields on the fibers of the $S^3$ bundle. But this is of course a restriction, and later on we will consider any kind of loops, which have a Fourier expansion 
\bea
C^{i}(s)=x^i+\sum_m \alpha^{i}_m e^{ims}
\eea
but for now we will restrict to the subspace of loop space consisting of points
\bea
(x^i,\alpha^{i}_n)=(0,\alpha^{i}_{+1},\alpha^{i}_{-1},0,0,...)
\eea
where
\bea
\alpha^{i}_{-1}&=&(\alpha,-i\alpha,\beta,-i\beta)\cr
\alpha^{i}_{+1}&=&(\bar{\alpha},i\bar{\alpha},\bar{\beta},i\bar{\beta}),\label{Hopfvar}
\eea
that is, to loops that are big circles on $S^3$. The Hopf map is given by
\bea
X+iY&=&2\alpha\bar{\beta}\cr
Z&=&\alpha\bar{\alpha}-\beta\bar{\beta}\label{Hopfmap}
\eea
where $X^I=(X,Y,Z)$ are real. Indices $I,J,...$ will be rised and lowered by $\delta_{IJ}$,  not by the metric that is induced by the Hopf map. We have that
\bea
R^2:=X^2+Y^2+Z^2=\(|\alpha|^2+|\beta|^2\)^2
\eea
so this is a projection from $S^3$ to $S^2$. We will denote the fiber over the point $X^I$ by $C_X$ and it is given by the loop $(e^{is}\alpha,e^{is}\beta)$.

Associated with the Hopf map we define the fields
\bea
A_{I}(X)&=&\partial_I \alpha^{in} A_{in}\cr
F_{IJ}(X)&=&\partial_I \alpha^{in} \partial_J \alpha^{jm} F_{in,jm}\cr
\phi(X)im\alpha^{im}&=&\phi^{im}
\eea
for $m,n=\pm 1$. Let us first consider the Abelian case and a static magnetically charged string (i.e. of Dirac type) in five space-dimensions. The projected gauge field $A_I(X)$ will then be that of a Dirac monopole in $3$ space dimensions. If we assume that the Abelian field strength produced by a monopole string is given by
\bea
H_{ijk}(x)&=&\epsilon_{ijkl}\frac{x^l}{|x|^4}
\eea
and associated loop field is defined as
\bea
F_{is,jt}(C)&:=&H_{ijk}(C(s))\dot{C}^k(s)2\pi\delta(s-t),
\eea
then the projected field strength will be given by
\bea
F_{IJ}(X)=2\epsilon_{IJK}\frac{X^K}{R^3}
\eea
This fact follows from a projection identity,\footnote{On the level of cohomology this result is closely related to the projection formula in \cite{Bott-Tu}. More generally this is related to what is called `integration along the fiber'. In \cite{Bott-Tu} an extensive presentation of the Hopf map can also be found.}
\bea
\frac{1}{2}\epsilon_{ijkl}\sum_{m,n,p,q}d\alpha^{i}_m\wedge d\alpha^{j}_n \alpha^{k}_p\alpha^{l}_q iq\delta_{m+n+p+q}=\epsilon_{IJK}\frac{1}{R}dX^I\wedge dX^J X^K
\eea
Here we should really take the pull-back of the left-hand side, that is, make the replacements $d\alpha^i_m=\partial_I \alpha^i_m dX^I$. If we define the Fourier transformed area elements
\bea
\sigma^{ij}_p=\sum_m im\alpha^{i}_m\alpha^{j}_{p-m}
\eea
then we find that only $\sigma^{ij}_{p=0}$ is non-zero (this is of course true only when $\alpha^{i,\pm 1}$ are the only non-zero components). Hence we can rewrite the projection identity in the follwing equivalent form,
\bea
\frac{1}{2}\epsilon_{ijkl}\sum_{m,n}d\alpha^{im}\wedge d\alpha^{j}_m \alpha^{kn}\alpha^l_n in=\epsilon_{IJK}\frac{1}{R}dX^I\wedge dX^J X^K
\eea
We derive this form of the projection identity by brute force in the appendix \ref{projid}. We can also write this in the form 
\bea
\epsilon_{ijkl}\sum_{m,n}d\alpha^{im}\wedge d\alpha^{jn} \alpha^k_m\alpha^l_n in=\epsilon_{IJK}\frac{1}{R}dX^I\wedge dX^J X^K\label{both}
\eea
because only $m+n=0$ can give a non-zero contribution, and there are only two such possibilities, either $m=1$, $n=-1$ or $m=-1$, $n=1$. Therefore we get the factor of $2$ in front.

It is now natural to examine what implication this projection has for the Bogomolny equation. Noting that
\bea
D^k_m\phi^l_n = in\delta_{m+n}\eta^{kl}\phi(X)+in\alpha^l_n(\partial^k_m X^I)D_I\phi(X)
\eea
we find that
\bea
F_{IJ}(X)=\epsilon_{ijkl}\partial_I\alpha^{im}\partial_J\alpha^{jn}in\alpha_n^l\underbrace{(\partial^k_m X^K)}_{\leadsto 2R(\partial^K\alpha^k_m)}D_K\phi(X)
\eea
which, if we can make the replacement as indicated in the underbrace, by noting the second projection identity (acting with the exterior derivative $d$ on both sides of (\ref{both}) produces yet another factor of $2$ in the left-hand side) becomes
\bea
F_{IJ}(X)=\epsilon_{IJK}D^K\phi(X).\label{Bog2}
\eea
This is the familiar Bogomolny equation in Yang-Mills-Higgs theory. We now have to show that we really can make that replacement. From $\alpha_{im}\alpha^{im}=2R$ it immediately follows that
\bea
\alpha_{im}X_I \(\partial^{im}X^I-2R \partial^I \alpha^{im}\)=0
\eea
but of course what we need is a finer identity than this (with less tensors being contracted). To verify this identity is highly technical so we have put this derivation in the appendix \ref{pecid}.

\section{Solitonic BPS string solutions}\label{bps}
We define the projections
\bea
(x,\alpha)\mapsto X^{I}_p:=\frac{1}{2}X^I_{ij}\sigma^{ij}_p(C)\label{map}
\eea
where
\bea
\sigma^{ij}_p(C)&:=&\int \frac{ds}{2\pi} e^{-ips} \dot{C}^i(s)C^j(s)\cr
&=&\sum_m im\alpha_m^i\alpha_{p-m}^j
\eea
is a Fourier transformed area element, and where the matrices $X^I_{ij}$ are certain constant anti-symmetric matrices. They are related to the Hopf map that projects $S^3$ to $S^2$ in a way that will be specified in a moment. Here we use a notation where $x^i$ is included as $\alpha^i_0$. For $p\neq 0$ we thus find that $X^I_p$ depends {\sl{linearly}} (as opposed to quadratically) on $x^i$. Also, for $p=0$ we find that $X^I_0$ does not depend on $x^i$ at all.

Instead of trying to solve (\ref{Bog1}) in loop space, we will consider a quotient space where we identify any two loops $(x,\alpha)$ and $(x',\alpha')$ which get projected to the same coordinates, $X^I_p(x,\alpha)=X^I_p(x',\alpha')$. On this quotient space we obtain a simpler Bogomolnyi equation to which one can find solutions by standard means.

As the notation suggests, we will let
\bea
(x,\alpha)\mapsto X^I_{p=0}=X^I
\eea
be the Hopf map when we restrict $(x,\alpha)$ as in Eq (\ref{Hopfvar}). Hence the anti-symmetric matrices $X^I_{ij}$ that we introduced above, sit in the Hopf map as follows,
\bea
X^I=\frac{1}{2}\sum_{m=\pm 1} X^I_{ij}\sigma^{ij}
\eea
where $\sigma^{ij}:=\sigma^{ij}_{p=0}$ is the usual anti-symmetric area element associated with the fiber over $X^I$. But this does not quite fix the matrices $X^I_{ij}$. The reason is that the area elements for the fibers are not all independent. From (\ref{Hopfvar}) we get the following relations,
\bea
\sigma^{13}&=&\sigma^{24}\cr
\sigma^{14}&=&-\sigma^{23}
\eea
For a certain reason that will become apparent later, we will choose the representation for the matrices where we have
\bea
X^I_{13}&=&X^I_{24}\cr
X^I_{14}&=&-X^I_{23}\label{relations1}
\eea
and also
\bea
X^I_{ij}=-X^I_{ji}\label{relations2}
\eea
With these additional prescriptions, these matrices are now uniquely determined by the Hopf map.

We now transform our loop fields $A_{im}$ and $\phi_{im}$ to the coordinates $X^I_p$ in quotient space according to the rules\footnote{Here the argument $X$ means $(X^I_p)$ where $I$ and $p$ run over all possible values.}
\bea
A_{im}(x,\alpha)&=&\partial_{im}X^{Ip}A_{Ip}(X)\cr
\phi_{im}(x,\alpha)&=&\sum_p i\(-m+\frac{p}{2}\)\alpha_{i,m-p}\phi_p(X)
\eea
and ask what implications the Bogomolnyi equation 
\bea
F_{im,jn}=\epsilon_{ijkl}D^k_{(m}\phi^l_{n)}
\eea
has on these new fields. The answer is that the new fields satisfy the simpler Bogomolnyi equation
\bea
F_{Ip,Jq}=\epsilon_{IJK}D^K_{(p}\phi_{q)}.\label{bog2}
\eea
provided that
\bea
\epsilon_{ijkl}iX^K_{kk'}+\epsilon_{ijkk'}iX^K_{kl}=\(X^I_{ik'}X^J_{jl}+X^I_{il}X^J_{jk'}\)\epsilon_{IJK}.\label{Hopf}
\eea
This equation is satisfied by the matrices $X^I_{ij}$ that we have given through Eqs (\ref{map}), (\ref{relations1}), (\ref{relations2}) as one may check by going though case by case. Another way to see this is by considering the `inverse projection identity'
\bea
\epsilon_{ijkl}in\alpha^l_n\partial^k_m X^K = \partial_{im}X^I \partial_{jn}X^J \epsilon_{IJK}.\label{master1}
\eea
where both sides are to be symmetrized in $(m,n)$. We show in appendix \ref{invid} that this equation is satisfied for $m,n=\pm 1$ by $X^I$ being the Hopf map. Had all the $\alpha^i_{\pm 1}$ been linealy independent, Eq (\ref{Hopf}) would have followed from this identity. But they are not as there are relations between them, as given through Eq (\ref{Hopfvar}). This implies that we should let the $X^I_{ij}$ be subject to the corresponding identifications in Eqs (\ref{relations1}), (\ref{relations2}), which then unambigously determine these matrices.

Solutions to Eq (\ref{bog2}) are easy to come by. If we define $Z=X_p+X_{-p}$, and put
\bea
\phi_m&=&(\delta_{m+p}+\delta_{m-p})\phi(Z)\cr
A_{Im}&=&(\delta_{m+p}+\delta_{m-p}) A_I(Z)
\eea
for any fixed non-zero\footnote{We could of course also take $p=0$, but in this case our solution would not depend on $x^i$, and could hardly be interpreted as a string solution! This would yield an interesting solution in loop space, but which would be trivial (constant) in spacetime.} integer number $p$, then we find that the Bogomolnyi equation reduces to 
\bea
F_{IJ}(Z)=\epsilon_{IJK}D^K\phi(Z)
\eea
which of course is a well-known equation, which can be solved by the Nahm procedure. The most famous solution to it is probably the Prasad-Sommerfield solution in the case of $SU(2)$ gauge group,
\bea
\phi^a(Z)&=&\frac{Z^a}{|Z|^2}H(v|Z|)\cr
A^a_I(Z)&=&\epsilon^{aIJ}\frac{Z_{J}}{|Z|^2}\(1-K(v|Z|)\)
\eea
where 
\bea
H(y)&=&y\coth y -1\cr
K(y)&=&\frac{y}{\sinh y}.
\eea
Hence $v=\sqrt{\phi^a(\infty)\phi^a(\infty)}$. 

We would like to think of $F_{IJ}(Z)$ as an element in the first Chern class. Nothing that we have said so far contradicts this assumption as we have said nothing about how $A_{im}$ should behave globally. 

One of our main ideas in this paper is that usual tensor multiplet fields -- the five scalar fields, a two-form gauge potential (with selfdual field strength), and the fermions -- are useful concepts only for $U(1)$ gauge group. If we for instance break $SU(2)$ gauge group down to $U(1)$, then it should make sense to speak about these tensor multiplet fields in an effective theory. It would be very interesting to derive this effective theory from our loop space theory. Our conjecture is that tensor multiplet fields in a $U(1)$ theory can be obtained from corresponding loop fields. We can for instance consider averages like
\bea
\<F_{\mu m,\nu n}(x,\alpha)\alpha_{\rho p}\> &=& i\<1\>\Lambda^2 h^{(m,n,p)}_{\mu\nu\rho}(x)+...\cr
\<\phi_{\mu m}(x,\alpha)\alpha_{\nu n}\> &=& i\<1\>\Lambda^2 \eta_{\mu\nu}\phi^{(m,n)}(x)+...
\eea
Here $\Lambda$ is the cut-off length of the loops. Apriori the dots could be other kinds of tensor fields. But these have to vanish identically since supersymmetry excludes any other fields than those that exist in the Abelian tensor multiplet. We can ensure this by imposing constraints such as that the loop fields be odd under orientation reversal of the loop. 

We would now like to argue that $h^{(m,n,p)}_{\mu\nu\rho}(x)$ and $\phi^{(m,n)}(x)$ for some values of $m,n,p$ (or some linear combination thereof) may be identified as the Abelian fields in the tensor multiplet.

We first consider a magnetically charged string of Dirac type,
\bea
F_{IJ}(Z)&=&\epsilon_{IJK}\frac{Z^K}{|Z|^3}\cr
\phi(Z)&=&\frac{1}{|Z|}+v.
\eea
(which is also be the asymptotic field configuration of a Prasad-Sommerfield string after that we have made a gauge rotation so that it points in one direction everywhere in internal space, which defines our $U(1)$ gauge group). We then find that (for some suitable choices of $(m,n,p)$),
\bea
h_{ijk}(x)&\sim& \epsilon_{IJK}X^I_{ii'}X^J_{ji'}X^K_{kl}\frac{x^l}{|x|^4}\cr
&\sim& \epsilon_{ijkl}\frac{x^l}{|x|^4}
\eea
We have then used that $F_{im,jn}=\partial_{im} X^{Ip}\partial_{jn}X^{Jq}F_{Ip,Jq}$ and the fact that $\epsilon_{IJK}X^I_{ii'}X^J_{ji'}X^K_{kl}\sim \epsilon_{ijkl}$ because this is the only invariant tensor of $SO(4)$ which is anti-symmetric in $ijk$. We have included the factor $1/|x|^4$ for dimensional reasons. In principle it should be possible to obtain this factor by a straightforward computation of the average, but we have not managed to do it.

For the scalar field we expect from dimensional analysis, to find the behaviour
\bea
\phi(x)=V+\frac{Q}{|x|^2}
\eea
for some specific values of $V$ and $Q$ that we have not managed to compute though. 

These are now precisely the behaviours one expects to find for the asymptotic Abelian tensor multiplet fields in the presence of a selfdual string at the origin.

\section{How to recover Abelian theory and determine $N_{\Lambda}$}\label{abelian}
Let us define the quantity 
\bea
b_{\mu\nu}(x):=\frac{1}{\<1\>}\sum_{m\neq 0}\frac{2m}{i\Lambda^2}\<A_{[\mu m}(x,\alpha)\alpha^m_{\nu]}\>\label{def}
\eea
Ideally we would like to compute correlation functions for $b$, or at least the partition function, on topologically non-trivial manifolds using the covariantized version of the flat  loop space action
\bea
\int d^6 x \frac{N_{\Lambda}}{4}\<F_{\mu m,\nu n}F^{\mu m,\nu n}\>.\label{loopaction}
\eea
(which presumably can be done by using the curved loop space metric $G_{\mu s,\nu t}(C)=G_{\mu\nu}(C(s))\delta(s-t)$). This seems to be a quite tough exercise though, so here we will content ourselves with just computing the propagator for $b$ on the topologically trivial flat space-time ${\mb R}^{1,5}$ using the action (\ref{loopaction}) and the definition (\ref{def}) of $b$ (and nothing more). We will find the same result as if we computed this propagator from the action
\bea
\int d^6 x \frac{1}{12}h_{\mu\nu\rho}h^{\mu\nu\rho}\label{spaceaction}
\eea
where $h=db$. Let us compute $h$. We find that
\bea
h_{\mu\nu\rho}(x)=\frac{1}{\<1\>}\sum_n \frac{2n}{i\Lambda^2}\<F_{[\mu 0,\nu |n|}(x,\alpha)\alpha^n_{\rho]}\>
\eea
Here we have used that
\bea
\<(\partial_{\nu n}A_{\mu 0})\alpha^n_{\rho}\>=\<\partial_{\nu n}\(A_{\mu 0}\alpha^n_{\rho}\)\>=O(\Lambda^{-1})
\eea
which vanishes (as $\Lambda\rightarrow \infty$) because for $n\neq 0$ we have a total derivative that we integrate over when taking the average. Now it is clear that $h$ is gauge invariant because $F$ is gauge invariant. But we are still far away from having showed that this $h$ behaves like an Abelian gauge field strength in all respects. 

We define the conjugate momentum of $b_{ij}(x)$ as
\bea
e^{kl}(y)&=&\frac{1}{N_{\Lambda}\<1\>}\sum_{m\neq 0}\frac{2m}{i\Lambda^2}\<E^{km}(y,\beta)\beta_m^l\>
\eea
The introduction of the factor $N_{\Lambda}$ is motivated by the fact that the action (\ref{loopaction}) is not canonically normalized. The conjugate momentum $E^{km}$ computed from that action is $N_{\Lambda}$ times the momentum one would get from a canonically normalized action (that is, the action (\ref{loopaction}) without the factor $N_{\Lambda}$). Therefore we have divided $E^{km}$ by $N_{\Lambda}$ and might then hope that $e^{ij}$ will turn out to be the conjugate momentum to $b_{ij}$ for the canonically normalized action (\ref{spaceaction}). Using the canonical commutation relations
\bea
[A_{im}(x,\alpha),E^{jn}(y,\beta)]=\delta_i^j\delta_m^n\delta^5(x-y)\prod_m\delta(\alpha_m-\beta_m)
\eea
we get the commutation relations,
\bea
[b_{ij}(x),e^{kl}(y)]&=&-\frac{\sum_{m,n}\frac{4mn}{\Lambda^4}\<\[A_{im}(x,\alpha),E^{kn}(y,\beta)\]\alpha^m_j\beta^l_n\>_{\alpha,\beta}}{\<1\>^2 N_{\Lambda}}\cr
&=&\frac{\<1\>_{\Lambda/\sqrt{2}}}{N_{\Lambda}\Lambda^2\<1\>_{\Lambda}^2}\delta_{ij}^{kl}\delta^5(x-y)
\eea
If we then notice that
\bea
\<1\>_{\Lambda}=\frac{1}{\Lambda^{6} 2^6 \pi^9}
\eea
we see that these commutation relations become canonical (i.e. with the right-hand side being unity) if we take
\bea
N_{\Lambda}=\frac{8}{\Lambda^2\<1\>}
\eea
which can also be expressed as
\bea
N_{\Lambda}=2^9\pi^9\Lambda^4.
\eea

In a free theory on topologically trivial Minkowski space essentially all observables can be constructed from the propagator. It does not follow from dimensional analysis that $\<b_{\mu\kappa}(p)b_{\nu\tau}(-p)\>\sim p^{-2}$ in momentum space, because the momentum is not the only dimensionful parameter of the problem -- we also have the dimensionful cut-off length $\Lambda$. It is even less obvious that we would get precisely
\bea
\<b_{\mu\nu}(p)b^{\kappa\tau}(-p)\>=\frac{2}{p^2}\delta_{\mu\nu}^{\kappa\tau}.
\eea
with the above choice of normalization factor $N_{\Lambda}$. 

We compute the left-hand side:
\bea
&&-\sum_{m,n}\frac{4mn}{\Lambda^4 \<1\>_{\alpha}\<1\>_{\beta}}\<\<A_{\mu m}(x,\alpha)A_{\nu n}(y,\beta)\>\alpha_{\kappa}^m\beta_{\tau}^n\>_{\alpha,\beta}
\eea
From the Abelian action (\ref{loopaction}) we get the gauge loop field propagator in Feynman gauge as
\bea
\<A_{\mu m}(x,\alpha)A_{\nu n}(y,\beta)\> = \frac{\eta_{\mu\nu}\delta_{m+n}}{N_{\Lambda}} 
\int \frac{d^6 p}{(2\pi)^2}\[\frac{d^6\pi}{(2\pi)^6}\]\frac{e^{-ip.(x-y)-i\pi.(\alpha-\beta)}}{p^2+\pi.\pi}
\eea
We now use that
\bea
\<e^{-i\pi.\alpha}\>_{\alpha}&=&\<1\>_{\alpha}e^{-\frac{\Lambda^2}{4}\sum_p \pi_p.\pi_{-p}p^{-2}}\cr
\<e^{-i\pi.\alpha}\alpha_{\mu m}\>_{\alpha}&=&\frac{1}{2}\<1\>_{\alpha}i\Lambda^2\pi_{\mu m}m^{-2}e^{-\frac{\Lambda^2}{4}\sum_p \pi_p.\pi_{-p}p^{-2}}
\eea
and 
\bea
\int d^6 Q \frac{1}{P^2+Q^2}e^{-\frac{\Lambda^2}{2m^2}Q^2} = 8V_5 m^6 \Lambda^{-6}P^{-2}+O(\Lambda^{-7})
\eea
to get the left-hand side as
\bea
&=&-\frac{1}{N_{\Lambda}}\eta_{\mu\nu}\eta_{\kappa\tau} \int \frac{d^6p}{(2\pi)^6} I
\eea
where
\bea
I\equiv -\frac{4}{6}\frac{\partial}{\partial(\Lambda^2)} \(\prod_{m\neq 0} \frac{8V_5 m^6\Lambda^{-6}}{(2\pi)^6}\)\frac{1}{p^2}
\eea
Using zeta function regularization 
\bea
\prod_{m\neq 0} a=\prod_{m>0}a^2&=&a^{-1}\cr
\prod_{m>0} m^2&=&2\pi,
\eea
we get
\bea
I=-\frac{12(2\pi)^{12}\Lambda^4}{8V_5 p^2}
\eea
Using that $V_5=\pi^3$, we get the left-hand side
\bea
&=&\frac{2^{10}\pi^9\Lambda^4}{N_{\Lambda}} \int \frac{d^6 p}{(2\pi)^6} e^{-ip.(x-y)} \frac{1}{p^2}
\eea
For this to become equal to the right-hand side, we should take
\bea
N_{\Lambda}=2^{9}\pi^9\Lambda^4.
\eea
This is the same value as we got earlier by other other means. We do not see any reason apriori why these two computations should yield the same answer. Of course this was necessary if we were to get a theory for an Abelian two-form. The fact that these two computations yield the same answer, we take as evidence for that the theory for a local Abelian two-form might be hidden in, or can be extracted from, our non-local Abelian loop space theory.

\section{Discussion}
There is an $A-D-E$ classification of the $(2,0)$ theories. The $A_r$ theories are realized in M-theory as the world-volume theories living on $r$ parallel M5 branes. If we separate one of the branes from the others, we get an Abelian tensor multiplet that interacts with massive loop fields. The separation amounts to giving the scalar field a vacuum expectation value $v$. We then expand the loop field $\phi_{\mu m}$ around this vacuum expectation value as
\bea
\phi_{\mu m}=v^a\sum_n in\alpha_{\mu n}\lambda^a_{m-n}+\phi'_{\mu m}
\eea
For the gauge field we then get a mass term
\bea
\tr\([A_{\mu m},\phi_{\nu n}][A^{\mu m},\phi^{\nu n}]\)=\mu^{ab,mn}A_{\mu m}^{a}A^{\mu,b}_n
\eea
with mass matrix
\bea
\mu^{ab,mn}=C^{ace}C^{bde}v^c v^d L_{m+n}(C)
\eea
where 
\bea
L_{m}(C)=\sum_n n(m-n)\alpha_{\mu n}\alpha^{\mu}_{m-n}=\int \frac{ds}{2\pi}e^{-ims}\dot{C}^2(s)
\eea
is the Fourier transformed length of the loop. In this sense $v$ gives the tension of W-boson string via the Higgs mechanism. 

It would now be interesting to compute the effective action for the Abelian tensor multiplet, obtained by integrating out the massive W-boson strings. We expect that this will produce the Hopf-Wess-Zumino term in \cite{Intriligator} that is needed for anomaly cancelation.

Another thing that could be interesting to check is whether the generalized Nahm equation for the self-dual string, that was proposed in \cite{Basu-Harvey}, can be related to the Nahm equation \cite{Nahm} via the Hopf map, in a similar fashion as we have related the Bogomolnyi equations to each other in this paper.

\newpage
\appendix
\section{Properties of the cut-off regularization}\label{cutoff}
To be slightly more general we consider a spacetime with one compact dimension $x^5$, around which loops may wind $w$ times,
\bea
C^5(s)&=&Rws + \sum_{n\neq 0} \alpha^5_n e^{ins}\cr
C^M(s)&=&x^{\mu} + \sum_{n\neq 0} \alpha^M_n e^{ins}
\eea
We then have
\bea
\<...\> &=& \sum_w e^{-\frac{1}{\epsilon^2}R^2w^2} \int [d^6\alpha] e^{-\frac{1}{\Lambda^2}\sum_p \alpha_p.\alpha_{-p}p^2}(...)
\eea
We define the generating functional
\bea
Z[J] &=& \sum_{{w}} e^{-\frac{2\pi}{\Lambda^2}R^2 {w}^2 + {w} R J_5^0}\int [d^6\alpha] e^{-\frac{1}{\Lambda^2}\sum_p \alpha_p.\alpha_{-p}p^2 +\sum_p i p\alpha_p.J_{-p}}
\eea
In the limit $R<<\Lambda$ we may treat ${w}$ as a continuos variable and the generated function be approximated by
\bea
Z[J]&=&Z[0] e^{\frac{\Lambda^2}{4}\sum_{p\in {\mb{Z}}} J_p^{\mu}J_{-p}^{\nu}\eta_{\mu\nu}}.
\eea
In the limit $R>>\Lambda$ only ${w}=0$ contributes, the higher winding loops being exponentially suppressed. In that limit the generating functional is approximated by
\bea
Z[J]&=&Z[0] e^{\frac{\Lambda^2}{4}\sum_{p\neq 0} J_p^{\mu}J_{-p}^{\nu}\eta_{\mu\nu}}.
\eea

Differentiating $Z[J]$ twice with respect to $J$ and then putting $J=0$, we find the propagators
\bea
\<\alpha_{m}^{\mu}\alpha_{n}^{\nu}\> &=& \frac{\Lambda^2}{2m^2}\delta_{m+n}\eta^{\mu\nu}\<1\>\cr
\<{w}^2\> &=& \left\{\begin{array}{ll}
\frac{\Lambda^2}{2 R^2}\<1\>, & R<<\Lambda\\
0, & R>>\Lambda
\end{array}\right.
\eea
As an application of this we have, for $R>>\epsilon$,
\bea
\<\dot{C}^{\mu}(s)\dot{C}^{\nu}(t)\>=\frac{\Lambda^2}{2}\(2\pi\delta(s-t)-1\)\<1\>.
\eea
We get any correlator $\<\bar{w}^n \alpha_{m_1}^{M_1}...\alpha_{m_r}^{M_r}\>$ by summing all Wick contractions. \footnote{Notice that the result presented in \cite{Nepomechie} is inconsistent. On the one hand we expect to have
\bea
\int ds \int dt \dot{C}^{\mu}(s)\dot{C}^{\nu}(t) = 0
\eea
for closed loops. Taking the average of zero, we should still get zero,
\bea
\int ds \int dt \<\dot{C}^{\mu}(s)\dot{C}^{\nu}(t)\> = 0
\eea
but if we plug in the result of \cite{Nepomechie}, we do not get zero.}

\section{Proof of the projection identity}\label{projid}
We find that
\bea
\alpha\bar{\alpha}&=&\frac{R+Z}{2}\cr
\beta\bar{\beta}&=&\frac{R-Z}{2}
\eea
but of course there is no unique way to express $\alpha$ and $\beta$ in terms of $X^I$. We have a one-parameter family of possible choices. If $(\alpha,\beta)$ is a solution, then the other solutions are obtained by letting $(\alpha,\beta)\rightarrow (e^{is}\alpha,e^{is}\beta)$. No restriction is made if we assume that $\alpha=\bar{\alpha}$ is real because we may always choose the parameter $s$ so that this is the case. Then we find that
\bea
\alpha&=&\frac{1}{\sqrt{2}}\sqrt{R+Z}\cr
\beta&=&\frac{X-iY}{2\alpha}
\eea
We compute the area elements
\bea
\sigma^{ij}:=\sum_m \alpha^{im}\alpha^j_m im
\eea
and find
\bea
\sigma^{12}&=&-2\alpha\bar{\alpha}=-(R+Z)\cr
\sigma^{13}=\sigma^{24}&=&-i\alpha(\beta-\bar{\beta})=-Y\cr
\sigma^{14}=-\sigma^{23}&=&-\alpha(\beta+\bar{\beta})=-X\cr
\sigma^{34}&=&-2\beta\bar{\beta}=-(R-Z)
\eea
and it is also easy to express these area elements in terms of $X^I$. We will also need the wedge products $d\alpha^i.d\alpha^j:=d\alpha^{in}\wedge d\alpha^j_n$,
\bea
d\alpha^1.d\alpha^2&=&0\cr
d\alpha^1.d\alpha^3=d\alpha^2.d\alpha^4&=&d\alpha\wedge d(\beta+\bar{\beta})\cr
d\alpha^1.d\alpha^4=-d\alpha^2.d\alpha^3&=&-d\alpha\wedge i d(\beta-\bar{\beta})\cr
d\alpha^3.d\alpha^4&=&2i d\beta\wedge d\bar{\beta}
\eea
From 
\bea
X&=&\alpha(\beta+\bar{\beta})\cr
Y&=&i\alpha(\beta-\bar{\beta})\cr
Z&=&\alpha^2-\beta\bar{\beta}
\eea 
we get
\bea
&&dX\wedge dY Z+dY\wedge dZ X +dZ\wedge dX Y \cr
&=& 2i(\alpha^2+\beta\bar{\beta})\(\alpha^2 d\beta\wedge d\bar{\beta}+\alpha\bar{\beta}d\alpha\wedge d\beta-\alpha\beta d\alpha\wedge d\bar{beta}\)
\eea
We then compute
\bea
\epsilon_{ijkl}d\alpha^i.d\alpha^j \sigma^{kl} = 8i\(-\alpha^2 d\beta\wedge d\bar{\beta} - \alpha\bar{\beta}d\alpha\wedge d\beta + \alpha\beta d\alpha\wedge d\bar{\beta}\)
\eea
Comparing these two expressions, we conclude that
\bea
\epsilon_{IJK}dX^I\wedge dX^J X^K = -\frac{R}{2} \epsilon_{ijkl}d\alpha^i.d\alpha^j \sigma^{kl}
\eea
This is the projection identity.

\section{A peculiar identity}\label{pecid}
We now wish to establish that
\bea
\partial_I \alpha^{[i|m|} \partial_m^{k]} X_J = \partial_I \alpha^{[i|m|} \partial_J \alpha^{k]}_m 2R
\eea
This appears to be about as weak as we can make this identity. It is for instance {\sl not} true that $\partial_m^k X_J=\partial_J \alpha^k_m 2R$.

Again we make the choice $\alpha=\bar{\alpha}$. We notice that, with our conventions,
\bea
\alpha^1.\alpha^3&=&X\cr
\alpha^2.\alpha^3&=&Y\cr
\alpha^1.\alpha^1&=&R+Z\cr
\alpha^3.\alpha^3&=&R-Z
\eea
We now consider the case that $X_I=X$, $X_J=Z$ and $i=1$, $k=3$ and compute the right-hand side in the above identity,
\bea
2R\(\partial_X \alpha^{1m} \partial_Z \alpha^3_m  - \partial_X\alpha^{3m} \partial_Z \alpha^1_m\) 
&=&2R \(\partial_X \alpha \partial_Z \bar{\beta} - \partial_X \beta \partial_Z \bar{\alpha} + c.c\) \cr
&=&2R \(\partial_X \frac{X+iY}{2\bar{\beta}} \partial_Z \bar{\beta} - \partial_Z\alpha\partial_X\frac{X+iY}{2\alpha} +c.c\)\cr
&=&\partial_Z\(\beta\bar{\beta}-\alpha^2\)\cr
&&+\frac{\alpha^2}{\bar{\beta}}\partial_Z\bar{\beta}-\frac{\beta\bar{\beta}}{\alpha}\partial_Z\alpha\cr
&&-\frac{2\alpha R}{\bar{\beta}}\partial_X \bar{\beta} \partial_Z\bar{\beta} 
  +\frac{2\bar{\beta} R}{\alpha}\partial_X \alpha \partial_Z \alpha +c.c.
\eea
We then notice that
\bea 
\frac{\partial_Z \beta}{\beta}=-\frac{\partial_Z \alpha}{\alpha}=\frac{\partial_Z \bar{\beta}}{\bar{\beta}}
\eea
and find that most terms cancel, leaving us with the right-hand side
\bea
=\partial_Z\(\beta\bar{\beta}-\alpha^2\)=\partial_Z (-Z)=-1.
\eea
We then compute the corresponding left-hand side,
\bea
\partial_X \alpha^{1m} \partial^3_m Z - \partial_X \alpha^{3m} \partial^1_m Z 
&=&-(\partial_X \alpha^{1m}) \alpha^3_m - (\partial_X \alpha^{3m}) \alpha^1_m \cr
&=&-\partial_X(  \alpha^{1m} \alpha^3_m)\cr
&=&-\partial_X X = -1
\eea
and thus we find agreement in this particular case.

Let us check another case. One for which the right-hand side is
\bea
\partial_X \alpha^{1m} \partial_Y \alpha^3_m  - \partial_X\alpha^{3m} \partial_Y \alpha^1_m 
&=& \partial_X \alpha \partial_Y \bar{\beta} - \partial_X \beta \partial_Y \bar{\alpha} +c.c\cr
&=& \partial_Y Z =0
\eea
Then the corresponding left-hand side is
\bea
\partial_X \alpha^{1m} \partial^3_m Y - \partial_X \alpha^{3m} \partial^1_m Y
&=&\partial_X \alpha^{1m} \alpha^2_m  - im\partial_X \alpha^{3m} \alpha^3_m \cr
&=&\partial_X \(im \(\alpha^{1m}\alpha^1_m - \alpha^{3m}\alpha^3_m\)\)\cr
&=&0
\eea
Again we find agreement. 

As one last check we check that the left-hand side is anti-symmetric under the exchange of $X$ and $Z$ (as the right-hand side is manifestly),
\bea
\partial_Z \alpha^{1m} \partial^3_m X - \partial_Z \alpha^{3m} \partial^1_m X
&=&(\partial_Z \alpha^{1m}) \alpha^1_m  - (\partial_Z \alpha^{3m}) \alpha^3_m \cr
&=&\partial_Z \frac{1}{2}\(\alpha^1.\alpha^1 - \alpha^3.\alpha^3\)\cr
&=&\partial_Z Z = 1.
\eea

We leave the other cases to check as exercises for the reader.

\section{The inverse projection identity}\label{invid}
Here we check that 
\bea
\epsilon_{ijkl}in\alpha^l_n\partial^k_m X_K = \partial_{im}X^I \partial_{jn}X^J \epsilon_{IJK}
\eea
holds when symmetrized in $(m,n)$. We notice that
\bea
\alpha^2_m&=&im\alpha^1_m\cr
\alpha^4_m&=&im\alpha^3_m\cr
\partial_{2m}&=&im\partial_{1m}\cr
\partial_{4m}&=&im\partial_{3m}
\eea
and that
\bea
X&=&\alpha^1.\alpha^3\cr
Y&=&\alpha^2.\alpha^3\cr
Z&=&\frac{1}{2}(\alpha^1.\alpha^1-\alpha^3.\alpha^3)
\eea
which yields
\bea
\partial_{im}X&=&(\alpha^3_m,im\alpha^3_m,\alpha^1_m,im\alpha^1_m)\cr
\partial_{im}Y&=&(-im\alpha^3_m,\alpha^3_m,im\alpha^1_m,-\alpha^1_m)\cr
\partial_{im}Z&=&(\alpha_m^1,im\alpha_m^1,-\alpha^3_m,-im\alpha_m^3)
\eea
We then compute
\bea
\partial_{1m}X\partial_{2n}Y-\partial_{1m}Y\partial_{2n}X&=&(1-mn)\alpha^3_m\alpha^3_n\cr
\epsilon_{1234}in(\alpha^4_n\partial^3_m Z-\alpha^3_n\partial^4_m Z)&=&(1-mn)\alpha^3_m\alpha^3_n
\eea
\bea
\partial_{1m}Y\partial_{2n}Z-\partial_{1m}Z\partial_{2n}Y&=&mn\alpha^3_m\alpha^1_n-\alpha^3_n\alpha^1_n\cr
\epsilon_{1234}in(\alpha^4_n\partial^3_m X-\alpha^3_n\partial^4_m X)&=&(mn-1)\alpha^3_n\alpha^1_n
\eea
\bea
\partial_{1m}Z\partial_{2n}X-\partial_{1m}X\partial_{2n}Z&=&in(\alpha^1_m\alpha^3_n-\alpha^1_n\alpha^3_m)\cr
\epsilon_{1234}in(\alpha^4_n\partial^3_m Y-\alpha^3_n\partial^4_m Y)&=&i(n-m)\alpha^1_m\alpha^3_n
\eea
We see that when symmetrized in $(m,n)$ these expressions become identical. The remaining cases can be worked out in a similar way.


\vskip 0.5truecm
\newpage


\begin{thebibliography}{999}
\bibitem{Witten}
  E.~Witten,
  ``Some comments on string dynamics,''
  arXiv:hep-th/9507121.
\bibitem{Hitchin}
  N.~J.~Hitchin,
  ``Lectures on special Lagrangian submanifolds,''
  arXiv:math.dg/9907034.
\bibitem{Teitelboim}
  C.~Teitelboim,
  ``Gauge invariance for extended objects,''
  Phys.\ Lett.\ B {\bf 167}, 63 (1986).\\
C. Teitelboim, `How commutators of constraints reflect the spacetime structure', Ann. Phys. 79:542 (1973).\\
P. A. M. Dirac, `The hamiltonian form of field dynamics', Canad. J. Math. 3 (1951), 1.
\bibitem{AG1}
   A.~Gustavsson,
  ``A reparametrization invariant surface ordering,''
  JHEP {\bf 0511}, 035 (2005)
  [arXiv:hep-th/0508243].
\bibitem{AG2}
  A.~Gustavsson,
  ``The non-Abelian tensor multiplet in loop space,''
  JHEP {\bf 0601}, 165 (2006)
  [arXiv:hep-th/0512341].
\bibitem{Nepomechie}G. Freund, R. Nepomechie, 
  ``Unified geometry of antisymmetric tensor gauge fields and gravity'', 
  Nucl. Phys. B199 (1982).
\bibitem{Alvarez}
  O.~Alvarez, L.~A.~Ferreira and J.~Sanchez Guillen,
  ``A new approach to integrable theories in any dimension,''
  Nucl.\ Phys.\ B {\bf 529}, 689 (1998)
  [arXiv:hep-th/9710147].
\bibitem{Schreiber}U.~Schreiber,
  ``From loop space mechanics to nonabelian strings,''
  arXiv:hep-th/0509163.
\bibitem{Akhmedov}
  E.~T.~Akhmedov,
  ``Towards the theory of non-Abelian tensor fields. I,''
  Theor.\ Math.\ Phys.\  {\bf 145}, 1646 (2005)
  [Teor.\ Mat.\ Fiz.\  {\bf 145}, 321 (2005)]
  [arXiv:hep-th/0503234].
\bibitem{Witten:1996hc}
  E.~Witten,
  ``Five-brane effective action in M-theory,''
  J.\ Geom.\ Phys.\  {\bf 22} (1997) 103
  [arXiv:hep-th/9610234].
\bibitem{Henningson:1999dm}
  M.~Henningson, B.~E.~W.~Nilsson and P.~Salomonson,
   ``Holomorphic factorization of correlation functions in  (4k+2)-dimensional
  (2k)-form gauge theory,''
  JHEP {\bf 9909}, 008 (1999)
  [arXiv:hep-th/9908107].
\bibitem{Bott-Tu}R. Bott, L. Tu, ``Differential forms in algebraic topology,'' (1982) Springer-Verlag, New York.
\bibitem{Intriligator}
  K.~A.~Intriligator,
   ``Anomaly matching and a Hopf-Wess-Zumino term in 6d, N = (2,0) field
  theories,''
  Nucl.\ Phys.\ B {\bf 581}, 257 (2000)
  [arXiv:hep-th/0001205].
\bibitem{Basu-Harvey}
  A.~Basu and J.~A.~Harvey,
  ``The M2-M5 brane system and a generalized Nahm's equation,''
  Nucl.\ Phys.\ B {\bf 713}, 136 (2005)
  [arXiv:hep-th/0412310].
\bibitem{Nahm}
  W.~Nahm,
  ``A simple formalism for the BPS monopole,''
  Phys.\ Lett.\ B {\bf 90}, 413 (1980).
\end{thebibliography}
\end{document}